

\font\tafont =  cmbx10 
\font\tbfont =  cmbx9

\magnification =1200

  \def\titlea#1#2{{~ \vskip 2 truecm 
  \tafont {\centerline  {#1}} 
  {\centerline{#2}}   } 
 \vskip 3truecm 
 }
 
\def \pn{\par\noindent}
\def \bs{\bigskip}
\def\titleb#1{\pn \bigskip \bigskip\parindent=0pt {\tbfont {#1}  } \bigskip
\parindent=16pt}
\def\Ref{\bs\pn{{\bf References}}\smallskip \parskip=5 pt\parindent=0pt}

\hfuzz=2pt
\tolerance=500
\abovedisplayskip=2 mm plus6pt minus 4pt
\belowdisplayskip=2 mm plus6pt minus 4pt
\abovedisplayshortskip=0mm plus6pt minus 2pt
\belowdisplayshortskip=2 mm plus4pt minus 4pt
\predisplaypenalty=0
\clubpenalty=10000
\widowpenalty=10000

\def \ka{\kappa}

\def \phi{\varphi}

\def \om{\omega}

\def \A{{\cal A}}

\def \D{{\cal D}}
\def \a{\alpha}

\def\grad{\nabla}

\def\({\left(}
\def\){\right)}

\def \d{{\rm d}}
\def\=#1{\bar #1}
\def\~#1{\widetilde #1}
\def\.#1{\dot #1}
\def\^#1{\widehat #1}
\def\"#1{\ddot #1}
\baselineskip 0.62 cm

\def \tr{transformation}
\def \com {constant of motion}
\def \coms {constants of motion}
\def \sy {symmetry}
\def \ss {symmetries}
{ \nopagenumbers

\titlea{Symmetries of dynamical systems} 
{and convergent normal forms}

\centerline{Giampaolo Cicogna}
\bigskip
\centerline{\it Dipartimento di Fisica, Universit\`a di Pisa}
\centerline{\it P.za Torricelli 2, I-56126 Pisa (Italy)}
\centerline{{\tt E-Mail: cicogna@ipifidpt.difi.unipi.it}}

\vskip 3cm
\pn
{\bf Abstract}. 
\pn
It is shown that, under suitable conditions, involving in 
particular the existence of analytic \coms , the presence of Lie point
symmetries can ensure the convergence of the transformation taking a 
vector field (or dynamical system) into normal form.

\vskip 5 truecm

\pn
{\tt PACS n. 03 20, 02 20 }

\vfill\eject}

The classical technique of transforming a given vector field \big(describing 
e.g. the flow of a dynamical system (DS)\big) into normal form (NF) 
(in the sense of Poincar\'e and Dulac) is a well known and useful method 
of investigation [1-2] (see also e.g. [3] for further Ref.); it is also 
well known, however, that Poincar\'e-Dulac 
series are in general only formal or asymptotic series. The 
convergence of a normalizing \tr\ is in fact a quite "rare" event, and 
often one considers truncated series \big(and then approximate \tr s 
(see e.g. [4-5] and Ref. therein)\big).  
In this note we want to propose shortly some result (details and complete 
proofs will be presented  in a separate paper) concerning the convergence 
of the normalizing \tr s: precisely, we will show that, under suitable 
conditions, the presence of some Lie \sy\ [6-7] of the vector field can ensure 
the convergence of the normalizing \tr . As a particular case, we recover  
a remarkable result which has been recently obtained by Bruno and Walcher        
for  2-dimensional problems [8]; in
the same context, see also [9] for an older result,  whose proof however has 
been recognized to be not complete [10].

We will consider $n-$dimensional vector fields $f$ in the form of DS
$$\dot u=f(u)=Au+F(u)\qquad \qquad u=u(t)\in R^n\eqno(1)$$
where $\dot u=\d u/\d t,\ f$ is assumed to be analytic in a neighbourhood 
of $u=0$, with $f(0)=0$, and
where the matrix $A\equiv(\nabla f)(0)$ is assumed to be nonzero and 
diagonalizable  (see [3] for a discussion on the  non-diagonalizable 
case). Introducing the usual notion of Lie-Poisson bracket 
$$\{f,g\}_k=(f\cdot\grad)g_k-(g\cdot\grad)f_k \qquad \qquad (k=1,\ldots,n)
            \eqno(2)$$
and that of "homological operator" $\A$ associated to the matrix $A$
$$\A(f)=\{Au,f\}=(Au)\cdot\grad f-Af \ ,\eqno(3)$$
it is known that a nonlinear vector function $h=h(u)$ is said to be in NF 
with respect to $A$ (or resonant with $A$) if
$$\A(h)=0\ . \eqno(4)$$
Let us also recall the following basic theorem [2]: 
A DS (1) can be put into NF by means of a convergent \tr\ if it fulfils 
the two following conditions:
\smallskip\pn 
{\sl Condition "A"}: there is a coordinate \tr\ changing $f$ to $\^f$, 
where $\^f$ has the form
$$\^f=Au+\a(u)Au$$
and $\a(u)$ is some scalar-valued power series  (with $\a(0)=0$);
\smallskip\pn
{\sl Condition "$\om$"}: let $\om_k=\min|(q,a)|$ (where $a_i$ are the 
eigenvalues of $A$, and parentheses stand for the scalar product) for all 
positive integers $q_i$ such that $\sum_{i=1}^n q_i<2^k$ and $(q,a)\ne 0$: 
then the series
$$\sum_{k=1}^\infty 2^{-k}\ln \om_k$$
is convergent.
\bs 
While condition $"\om"$ is weak condition, controlling the 
appearance of small divisors [2], condition "A" is clearly a rather strong 
restriction.  We explicitly assume that all the DSs considered in the remaining 
of this paper will satisfy condition $"\om"$, but do {\it not} satisfy 
condition "A". 

A vector function  
$$g(u)=Bu+G(u)\eqno(5)$$
(not proportional to $f$) is said to be a Lie point (time independent) 
\sy\ for the DS (1) if
$$\{f,g\}=0\ .\eqno(6)$$
In terms of Lie algebras, one says that the vector field 
operator $g\cdot\grad$  generates a symmetry of the DS. 

We now can state our results.
\pn
{\bf Theorem 1}. Assume that: $i$) the DS (1) admits an analytic \sy\ (5) 
where either the matrix $B$ is proportional to $A$, or $B=0$ and $G(u)$ 
is not proportional to $F(u)$; 
\pn 
$ii$) once in NF, the DS takes the form 
$$\.u=h(u) = Au+\a(u)Au+\mu(u)Mu \eqno(7)$$
where $M$ is some matrix (not proportional to $A$), $\a$ and $\mu$ some 
scalar functions,  and the two linear 
problems $\.u=Au$ and $\.u=Mu$ do not admit time-independent common \coms .
\pn
Then, the DS can be put in NF by means of a convergent normalizing \tr .
\smallskip\pn
{\it Sketch of the proof}. If $B=0$, consider the new \sy\ $g+f=Au+F+G$. 
Thanks to (4), the linear field $g_A=Au$ is a \sy\ for the NF. 
Using hypothesis $ii)$, one can show (see [11]) 
that the NF (7) does not admit \coms\ (i.e. functions $\ka=\ka(u)$ 
expressed by (possibly formal) series such that 
$h\cdot\grad\ka=0$); as a consequence, its only \sy\ (including 
nonlinear and possibly formal ones) having $Au$ as linear part 
is just $g_A=Au$ (let us recall that multiplying 
a \sy\ by a \com\ one obtains another \sy ). Then the coordinate \tr\ 
taking (1) into (7) transforms the \sy\ $g$ according to
$$g=Au+G\to g_A=Au$$
This means that condition "A" is satisfied by this \tr\ of the \sy : thus 
there is a normalizing \tr\ which is 
convergent, and one can easily conclude that under this 
\tr\ the DS  also is transformed into NF.
\medskip
\pn   
{\bf Theorem $2$}. Instead of $i)$  in Theorem 1 assume that the DS (1) 
admits  $\ell$ ($\ge 1$) analytic symmetries $g_j=B_ju+G_j(u)$, where  
the matrices $B_j(\not= 0)$  are linearly independent (and such that 
no linear combination is proportional to $f$), and
where $\ell$ is precisely the number of the linearly 
independent linear symmetries admitted by the DS once in NF. Then, with 
the condition $ii)$ as in Theorem 1, the same conclusion holds.
\smallskip\pn
{\it Sketch of the proof}. According to a general property of NFs (see 
[3,12]), the linear fields $B_ju$ are (linear) \ss\ for the NF. 
Together with $Au$, we have then $\ell+1$ \ss\ for the NF; 
therefore, there must be one \sy\ having just $Au$ as linear part, and 
the argument proceeds along similar lines as in Theorem 1.
\medskip

Notice that it can be easily seen that if the DS has dimension $n=2$, the 
assumption $ii)$ in the above Theorems is automatically satisfied, and 
then one reobtains the Bruno and Walcher result, namely:

\smallskip\pn
{\bf Corollary} [8]. If a $2-$dimensional DS admits an analytic \sy , it can 
be normalized by a convergent \tr .
\bs
We now give a quite general example, which can be interesting  both for 
its possible applications to Hamiltonian DSs, and for illustrating the 
role played by the presence of \ss\ in ensuring the convergence 
of the normalizing \tr s.

Let $n=2m$ and, putting $u\equiv(x_1,\ldots, x_m,y_1,\ldots,y_m)\in R^{2m}$; 
assume that a Lie group $\Gamma$ acts "diagonally" on both the 
$m$-dimensional spaces of the vectors $x$ and $y$ through the same linear 
representation $\D$: i.e. $x\to x'=\D x$, $y\to y'=\D y$, where $\D$ is an 
absolutely irreducible representation (i.e. the only matrix commuting 
with $\D$ is a multiple of the identity). Consider then a DS of 
the following form
$$\.u= f(u)=Au+F(u)\eqno(8)$$ 
where
$$ A =\pmatrix{0 & I_m \cr
               -I_m & 0 } \eqno(8')$$
and assume that $F(u)$ admits the symmetries $B_iu$, where $B_i$ are the 
(matrix representatives in the direct sum $\D\oplus\D$
of the) Lie generators of this group $\Gamma$ \big(the linear part $Au$ 
fulfils this symmetry requirement, so that the full DS 
(8) admits this symmetry\big). E.g. if the DS is
$$\.u=Au+p(u)u\eqno(9)$$
the scalar function $p$ must depend only on the quantities 
$\rho_a=\rho_a(u)$ which are invariant under $\D\oplus\D$ (e.g., in 
the case $m=3$, $\Gamma=SO(3)$ and $\D$ its fundamental 
representation, these are given by $x^2=(x,x)$, 
$y^2=(y,y)$, $x\cdot y=(x,y)$ where the parentheses stand for the scalar 
product in $R^3$). The NF of the DS (8) or (9) must admit the linear \ss\ 
$B_iu$ and $Au$ (see [3,12]), then it is easy to see that it must take 
the form
$$\.u=Au+\a Au+\mu u\eqno(10)$$
where $\a$ and $\mu$ are functions of the quantities invariant under all 
these \ss\ (e.g., of $r^2=x^2+y^2$ only, in the $SO(3)$ case). 
Assume now (cf. the example in [9])
that in the DS (9) the function $p$ is a 
homogeneous polynomial of degree $2k$ built up with the quantities $\rho_a$: 
it is easy to verify that the following vector function (with vanishing 
linear part)
$$g=(x^2+y^2)^k u\eqno(11)$$
generates a nontrivial analytic symmetry for the  DS (9), and that  
there are no common \coms\ for (10), as requested by $ii)$ in the above 
Theorems. Therefore, the convergence of the normalizing \tr\ is ensured 
by Theorem 1. It is important 
to notice that, if our problem would  not possess the \sy\
$\Gamma$, the NF (10) would contain many other terms in its r.h.s., and  
that it is precisely the presence of the symmetry 
which forces the NF to contain only  $A$ and the identity, and therefore 
allows us to apply the argument concerning the \coms .
This seems to confirm the conjecture [8,13] 
that the presence of a "sufficient" number of symmetries may be an essential 
request in order to guarantee the convergence of a normalizing \tr .

\titleb{Acknowledgments}

I am grateful to prof. A.D. Bruno for a very useful and clarifying 
discussion, and for his kind interest in this argument. Prof. Bruno 
informed me that he succeeded in extending, under suitable conditions,  
the results in [8] to the case of dimension $n=3$. I am also indebted to 
dr. Giuseppe Gaeta for useful suggestions. 
     
\vfill\eject

\Ref
[1] Arnold V.I., 1988, "Geometrical methods in the theory of
differential  equations", Springer, Berlin 

[2] Bruno A.D., 1989, "Local methods in nonlinear differential equations",
Springer, Berlin  

[3] Cicogna G. and  Gaeta G., 1994,  Journ. Phys. A: Math. Gen. 
{\bf 27}, 7115-7124

[4]  Bazzani A., Todesco E., Turchetti G., and Servizi G., 1994, "A normal 
form approach to the theory of nonlinear betatronic motion", 
CERN Report 94-02, Geneva 

[5] Cicogna and Gaeta G., Approximate symmetries in dynamical 
systems, Nuovo Cimento B, to be published

[6]  Olver P.J., 1986, "Applications of Lie groups to differential
equations", Springer, Berlin

[7] Ovsjannikov L.V., 1962, "Group properties of differential equations",
Novosibirsk; (English transl. by Bluman G.W., 1967); and 1982,
"Group analysis of differential equations", Academic Press, New York

[8] Bruno A.D. and Walcher S., 1994,
 J. Math. Anal. Appl. {\bf 183}, 571-576
 
[9]  Markhashov L.M., 1974, J. Appl. Math. Mech. {\bf 38},  788-790

[10] Bruno A.D., 1993 , Selecta Math. 
(formerly Sovietica) {\bf 12}, 13-23

[11] Walcher S., 1991, Math. Ann. {\bf 291} , 293-314

[12] Elphick C., Tirapegui E., Brachet M. E., Coullet P., and  
Iooss G., 1987, Physica {\bf D, 29}, 95

[13] Ito H., 1989, Comm. Math. Helv. {\bf 64}, 412-461; and 1992, Math. Ann. 
{\bf 292}, 411-444

\bye